\documentclass[conference]{IEEEtran}
\usepackage{amssymb, amsmath,graphicx,charter, latexsym}
\usepackage{amsthm}

\usepackage{amssymb,graphicx,amsmath,amsthm,multirow,url,float,epstopdf,cite}

\def\bs{\ensuremath\boldsymbol}
\begin{document}
\title{A Theory for the Operation of the Independent System Operator in a Smart Grid with Stochastic Renewables,  Demand Response and Storage}
\author{Rahul Singh, Ke Ma, Anupam Thatte, P.R. Kumar, Le Xie
\thanks{This material is based upon work partially supported by NSF under Contract Nos. CPS-1239116 and Science \& Technology Center Grant CCF-0939370.}
\thanks{The authors are with Department of Electrical and Computer Engineering, Texas A\&M University
       {\tt\small \{rsingh1,ke.ma\}@tamu.edu, anupam.thatte@gmail.com, prk@tamu.edu, lxie@ece.tamu.edu}}
 }

\maketitle
\IEEEpeerreviewmaketitle

\begin{abstract}

In this paper, we address a key issue of designing architectures and algorithms which generate optimal demand response in a decentralized manner for a smart-grid consisting of several stochastic renewables and dynamic loads. By \emph{optimal demand response}, we refer to the demand response which maximizes the utility of the agents connected to the smart-grid. By \emph{decentralized} we refer to the desirable case where neither the independent system operator (ISO) needs to know the dynamics/utilities of the agents, nor do the agents need to have a knowledge of the dynamics/utilities of other agents connected to the grid. The communication between the ISO and agents is restricted to the ISO announcing a pricing policy and the agents responding with their energy generation/consumption bids in response to the pricing policy.

We provide a complete solution for both the deterministic and stochastic cases. It features a price iteration scheme that results in optimality of social welfare. We also provide an optimal solution for the case where there is a common randomness affecting and observed by all agents. This solution can be computationally complex, and we pose approximations. For the more general partially observed randomness case, we exhibit a relaxation that significantly reduces complexity. We also provide an approximation strategy that leads to a model predictive control (MPC) approach. Simulation results comparing the resulting optimal demand response with the existing architectures
employed by the ISO illustrate the benefit in social welfare utility realized by our scheme. To the best of the authors' knowledge, this is the first work of its kind to explicitly mark out the optimal response of dynamic demand.
\end{abstract}

\section{INTRODUCTION}

Traditionally, given demand (or a demand forecast), generation (or the planned generation) has been dispatched so as
to balance demand in power systems. Since there are many generators capable of producing power at different cost curves, it is desirable to allocate the total power generation among the generators so that the total cost of generating the required power is minimized. This role of determining which generators are selected, and how much power they produce, has been played by the Independent System Operator (ISO).

This economic dispatch of generators is done by soliciting production bids (power vs. price curves) from each generator, and then, given the demand, choosing the generators such that the overall cost to purchase the power is minimized. Motivated by the problem of integrating renewable power generation from sources such as wind and solar \cite{Xie2011windintegration}, we consider the problem of demand response, i.e., adjusting demand so that it can be part of the flexibility to match the intermittent generation \cite{Rahimi2010}. The process of increasing or decreasing demand can be accomplished by decreasing or increasing the price of power. Thus the problem becomes one of choosing the level of demand as well as allocating the required generation among the various producers of power from fossil fuel sources such as coal or gas.

An additional consideration is that both power generators as well as loads are dynamic systems. Generators, for
example, have ramping constraints as well as maximum power constraints. Similarly, analysis of load data shows that loads also have certain dynamic constraints \cite{An2015}. Thus, all variables, including price as well as generator power outputs as well as loads, are \emph{functions of time}.

A third consideration is that renewable power production is uncertain, which we model as a stochastic process.
Thus, in addressing the dynamic evolution of the future demand as well as future non-renewable power generation
over time, one needs to take future uncertainty of renewable power generation into account.

In this paper, we consider the resulting overall problem faced by the ISO: Given stochastic renewable generation
which is disclosed causally in time, how should an ISO choose the price causally as a function of time, and thereby the
level of demand response elicited, and then allocate the net remaining generation among various conventional generators. A further consideration which we will neglect in this paper is that the network constraints are satisfied. However, this can be incorporated into the setup that we develop in this paper.

There are some complexities that are involved in the problem, and some constraints that a desirable solution has to
satisfy. It is desirable that the communication from the ISO to the generators or loads contain only the price of power in each market interval. Also it would be desirable that the communication from each generator and load to the ISO, at each time instant only be the resulting power-price point for desired production and demand response respectively. In particular, we would not like to require that each load provide a dynamic model of its behavior as well as its utility function to the ISO, and similarly for each generator. Not only is this a lot of information that the agents may not want disclose for reasons of privacy or business, but it also implies that the ISO solve a gigantic optimization problem encompassing every single agent in the system with all agents' utility functions.

Thus we would like the system-wide optimization be conducted by the agents themselves in a distributed manner,
coupled only by the price announced by the ISO. Additional complexities are introduced by the fact that the agents are dynamic systems as well as the fact that the renewable power generation is stochastic. Our approach attempts to provide a comprehensive solution that takes into consideration all individual agents' utility functions. It
has one feature that would be desirable to eliminate. At each time our approach requires that the ISO interact with the agents in an extended transaction. The ISO announces a tentative price, and the agents respond with power generation and consumption levels. Based on these responses, the ISO again announces a tentative price, and so on. This process continues till the price converges. We see no way to eliminate this bidding process, and wonder
if it is well nigh impossible in general since the agents need to somehow convey to the other agents how they respond to prices. At any rate, this is a significant open problem.

We examine a sequence of models, a deterministic model, a common completely observed randomness model, and a partially observed randomness model. For the deterministic model our solution is complete and leads to social welfare optimality. Next we consider the case where there is a common randomness affecting all agents and observed by each of them. This of course includes the case where only certain portions of the overall randomness affect individual agents and in different ways. In this case we propose a complete solution, in principle, which leads to social welfare optimality of the utility. In this case, we also propose a scheme to reduce complexity that leads to a model predictive approach. In the most general case where the different agents experience differing randomness not observable by other agents, we propose a relaxation that leads to a significant reduction in complexity. We also show that our MPC approach extends to this case. We also represent simulations comparing our algorithms with the approach used presently by the ISO and illustrate the improvement in utility that can be realized.

Overall, our approach provides a theory for the operation of the ISO in an environment where it is needed to integrate stochastic renewables, demand response, and dynamic and other constraints as well as uncertainties in both generation and loads. The solution requires communication only of prices and energy production/consumption responses, as is desirable. Specifically it is not necessary for the ISO to be aware of the dynamics of producers or consumers or their individual utility functions. Similarly, the various agents need only know their own dynamic system models and states and utility function, and need not be aware of any attributes of other users, or even the existence of other users. We hope our approach leads to a firm foundation for the operation of the next generation ISO.

One approach to procure Demand Response is direct load control, where based on an agreement between customers and the aggregator or utility, remote control of certain devices (e.g., air conditioners, pumps etc.) is used to manage their energy consumption \cite{Callaway2011}. The lack of user privacy is a major barrier to large scale implementation. The alternative is an incentive based approach such as real-time pricing where users voluntarily manage their consumption in response to a time-varying retail tariff \cite{Conejo2010}. The challenge with this approach is the difficulty faced by customers to manually respond to time-varying prices. We envision a scenario where an intelligent agent automatically manages the energy consumption scheduling of the customer based on certain cost and comfort settings selected by the user. The key issue in this case is of designing an architecture which yields optimal demand response in a decentralized manner while maintaining user privacy. From the system operator point of view the challenge is to optimally balance the system when uncertainty arises both in supply and demand, without resorting to the brute force and expensive option of procuring large amounts of reserve and/or energy storage \cite{Makarov2009}.

We realize that the current smart grid suffers from the key issues of renewable penetration and electricity price
fluctuations. Our approach of modeling the users by a dynamical system in lieu of following a ''static-optimization''
is the key to generate demand response that plays the dual role of mitigating renewable penetration and price fluctuation via utility optimization. Thus our goal is to provide a framework in which Demand Response could participate in both energy and ancillary service markets \cite{Koch2011, Galus2011}. Of course, this demand response needs to be optimized in order to achieve our pre-set goals. This is done via utilization of the computational power and latent energy storage that is present in the \emph{smart} users connected to the smart grid. 

The paper is organized as follows: we begin with a survey of some related works in Section II and give a
complete description of the problem in Section III. This is followed by a discussion on iterative bidding schemes and
the ensuing optimal demand response in a deterministic setting (Section IV) and stochastic setting (Section V).
Finally simulation results are shown (Section VI), which support our theory.

\section{LITERATURE SURVEY}
There have been many efforts since the deregulation of the electricity sector on a market-based framework to clear the system. Ilic et al. \cite{Ilic2011framework} proposed a two-layered approach that internalizes individual constraints of market participants while allowing the ISO to manage the spatial complexity. The approximated MPC algorithm is shown to work in many realistic cases.
 
In order to analyze the strategic interactions between the ISO and market participants, game theoretical approaches have been proposed in a number of paper. Zhu et al. \cite{Zhu2013} use a Stackelberg game framework for economic dispatch with demand response. The approach uses a two person game with ISO as leader and users aggregated into second player. The users change their demand based on price signal so as to maximize their payoff function. The ED problem considered is a single time interval conventional dispatch without transmission line constraints. Bu and Yu \cite{Bu2013game} models the interactions between electricity retailers and customers as a Stackelberg game. This work considers the case of a monopoly retailer where information about customers' utility and consumption pattern is available. Jia and Tong \cite{Jia2013} uses Stackelberg to study the energy consumption scheduling problem for customers who are subjected to a time-varying price which is determined one day ahead of time. The trade-off between consumer surplus and retailer profit under different pricing schemes is investigated.

Song et al. \cite{Song2000optimal} applies a Markov decision process (MDP) model to the bidding problem for generators participating in electricity market. Gajjar et al. \cite{Gajjar2003application} extends this approach and uses actor-critic learning. Gao et al. \cite{Gao2015optimal} presents a method for obtaining the bidding strategy of market participants using parametric linear programming. However, it assumes that market participants have complete information on system conditions and competitor strategies.

Wang et al. \cite{Wang2014gametheoretic} formulates the trading of energy by storage units as a noncooperative game. Under certain assumptions for the strategy space and utility functions a Nash equilibrium is shown to exist. An iterative algorithm is used to reach equilibrium following which a double auction is conducted. Mohsenian-Rad et al. \cite{Mohsenian2010} proposes a distributed algorithm to obtain the optimal energy consumption schedule for each user. The problem of determining the user energy consumption schedule for the whole day is formulated as a deterministic linear program. Two problems are considered with two different objectives: (i) minimize energy cost, and (ii) minimize the peak to average ratio of demand.

One of the major challenges in the above approaches is how to elicit optimal demand response without revealing the inherent dynamic nature of the loads to the ISO. In this paper, we model the users as stochastic dynamical
systems and generate the optimal demand response in a decentralized and adaptive manner, thus maximizing the
sum total of the utilities of the users, which in-turn allows for maximum renewable penetration and in controlling
price fluctuations. 
\section{Notation}
Throughout the paper $\omega,\omega_i$ etc. will be used to denote random variables. Also, random variables will be in capitals, while their realizations in small letters, eg. random variable $X$, and event $\{X=x\}$, etc.

\section{Problem Formulation}\label{formulation}
We consider a smart-grid in which there are a total of $N$ agents.
Each agent may be either a consumer or a producer of electricity. We model time as consisting of discrete periods. At each discrete time $t$ each agent $i$ obtains or supplies $u_i(t)$ units of energy (equivalently power since it is proportional to it given the fixed period) to the grid, with $u_i(t)>0$ signifying that user $i$ supplies energy to the grid at time $t$, while $u_i(t)<0$ signifying an energy consumption from the grid by the $i$-th user. We will suppose that there is net energy balance at each time over the whole grid: $\sum_{i=1}^N u_i(t) = 0$ for all $t$. This model does allow for storage too, since each storage device can be considered as an agent.

We model each agent as a dynamic system. The motivation in the case of an agent which is a generator is that it has
ramp up constraints, thus necessitating a dynamic system model, or in the case of a load it may have similar ramp
down constraints as well as delay in demand response. The state of the user $i$ at time denoted by $x_i(t)\in \mathcal{X}_i$ evolves as,
\begin{align}\label{dyn}
x_i(t+1) = f^t_i(x_i(t),u_i(t)), t = 1,2,\ldots,T-1.
\end{align}
Thus the state of the grid resides on the space $\otimes \mathcal{X}_i$.
We suppose that each agent $i$ has a utility function $F_i(\cdot):\mathcal{X}_i\mapsto \mathbb{R}$, with the understanding that the user prefers a state having higher utility. The total utility of user $i$ over the horizon $\{1,2,\ldots,T\}$ is $\sum_{t=1}^{T}F_i(x_i(t))$. (The theory can be generalized in a straightforward way to utilities that are time-dependent.)
 The model~\eqref{dyn} can incorporate constraints on inputs, for example reflecting bounds on ramp rates, such as $u_i(t) \in \mathcal{U}_i$. In that case, these constraint sets $\mathcal{U}_i$ are not dualized, but simply carry over to the dual in~\eqref{eq:4}. For simplicity of exposition we will not explicitly consider this case in the treatment here, but will consider such constraints in the numerical examples in Section~\ref{simulations}.

With the above set-up, we are led to the following deterministic social welfare optimization problem (DSWOP):
\begin{align}\label{p1}
 & \max \sum_{i=1}^{N}  \sum_{t=1}^{T} F_i(x_i(t))\notag\\
\mbox{ subject to } &\sum_i u_i(t) = 0, \forall t = 1,2,\ldots,T\notag\\
& x_i(t+1) = f^t_i(x_i(t),u_i(t)),\mbox{ for}\notag\\
& t = 1,2,\ldots,T-1, i=1,2,\ldots,N.
\tag{DSWOP}
\end{align}
Subsequently we will consider the stochastic version of the problem caused by uncertainties due to weather, etc.

\section{Optimal Demand Response and Decentralized Solution Via Bidding}\label{deter}
The above problem can be interpreted as giving the ISO the task of determining the $T$-dimensional vectors $\bs{u}_i:=\left(u_i(1),u_i(2),\ldots,u_i(T)\right)$, for $i=1,2,\ldots,N$, so as to maximize the social welfare $\sum_{i=1}^{N} \sum_{t=1}^{T}  F_i(x_i(t))$. In this section, we will derive an
easy-to-implement algorithm that does so, while satisfying information and action decentralization, with all
communication between agents restricted simply to being either price announcements or purchasing or supply of
energy decisions in response to prices. The ISO simply determines the appropriate prices causally, while each user
optimizes its response causally.

The Lagrangian for the problem~\ref{p1} is,
\begin{align}
&\mathcal{L}(\bs{u}_1,\bs{u}_2,\ldots,\bs{u}_N,\bs{\lambda})\notag\\
&:=\sum_{i=1}^{N}  \sum_{t=1}^{T} F_i(x_i(t))- \sum_ {t=1}^{T}\lambda(t)\left(\sum_{i=1}^{N} u_i(t)\right),
\end{align}
where $\lambda(t), t=1,2,\ldots,T$ are the Lagrangian multipliers associated with the constraints $\sum_i U_i(t)=0,t=1,2,\ldots,T$ respectively. The Lagrange dual function is,
\begin{align}\label{eq:4}
& D(\bs{\lambda}) = \max_{\bs{u}_1,\bs{u}_2,\ldots,\bs{u}_N} \mathcal{L}(\bs{u}_1,\bs{u}_2,\ldots,\bs{u}_N,\bs{\lambda})\notag\\
&=\max_{\bs{u}_1,\bs{u}_2,\ldots,\bs{u}_N} \sum_{i=1}^{N} \left( \sum_{t=1}^{T} F_i(x_i(t))- \lambda(t) u_i(t)\right),
\end{align}
The objective function~\eqref{eq:4} can be decomposed agent by agent since they are only coupled by price. Hence we
consider the optimal problem for agent $i$ as one of maximizing the objective
\begin{align}\label{eq:1}
 \max_{\bs{u}_i}  \sum_{t=1}^{T} F_i(x_i(t))- \lambda(t) u_i(t).
\end{align}
for the dynamic system~\eqref{dyn}. The optimal cost is a function of the initial condition and the Lagrange multiplier sequence
$\bs{\lambda} = \left(\lambda(1),\ldots,\lambda(T)\right)$., and we denote it $V_i(x_i(0),\bs{\lambda})$. Therefore,
\begin{align*}
D(\bs{\lambda}) = \sum_{i=1}^N V_i(x_i(0),\bs{\lambda}).
\end{align*}
We thus observe that the consideration of the dual problem has led us to a decentralized problem. Its solution involves the ISO first announcing the price vector $\bs{\lambda}$, and then each agent $i$ simply optimizing its own objective~\eqref{eq:1} by determining its vector $\bs{u}_i$. Thus neither the ISO, nor the other agents need to know the utility function of agent $i$. 
The dual problem is to,
\begin{align}\label{dual}
&\min D(\bs{\lambda}) \qquad\mbox{subject to}\notag\\
&\lambda(1),\ldots,\lambda(T) \geq 0.
\end{align}
We will suppose that strong duality holds, i.e., the optimal values of~\ref{p1} and~\eqref{dual} are equal. There are several sufficient conditions for strong duality.
For example a sufficient condition is for the utility functions $ \sum_{t=1}^{T} F_i(x_i(t))$ to be convex and the feasibility region of the problem~\ref{p1} to be non-empty. Denoting the optimal solution of the Dual problem by $\bs{\lambda}^{\star}$, we will suppose that,
\begin{align*}
& D(\bs{\lambda}^{\star}) = \sum_{i=1}^{N} V_i(x_i(0),\bs{\lambda}^{\star})\\
&= \max_{\substack{\bs{u}_i,i=1,2,\ldots,N:\\ \sum_i u_i(t) = 0,\\ \forall t = 1,2,\ldots,T}} \sum_{i=1}^{N}  \sum_{t=1}^{T} F_i(x_i(t)).
\end{align*}
The issue faced by the ISO is how to determine the optimal price vector $\bs{\lambda}^{\star}$.
Since $D(\bs{\lambda})$ as well as $V_i(x_i(0), \bs{\lambda})$ are all concave functions of $\bs{\lambda}$, will consider the use of the sub-gradient method for iterating on the price-vector $\bs{\lambda}$ so as to converge to the optimal price-vector $\bs{\lambda}^{\star}$. Denoting the sub-gradient by $\frac{\partial D}{\partial \bs{\lambda}}$, we note that,
\begin{align*}
&\frac{\partial D}{\partial \bs{\lambda}} = \sum_{i=1}^{N} \frac{\partial V_i}{\partial \bs{\lambda}}\\
&= \left(\sum_{i=1}^{N}u^{\lambda}_i (1),\sum_{i=1}^{N}u^{\lambda}_i (2),\ldots,\sum_{i=1}^{N}u^{\lambda}_i (T)\right),
\end{align*}
where $\bs{u}^{\bs{\lambda}}_i : =\left(u^{\lambda}_i (1),u^{\lambda}_i (2),\ldots,u^{\lambda}_i (T)\right)$ is the vector that achieves the optimal utility for the $i$-th user for the price vector $\bs{\lambda}$ in~\eqref{eq:1}.

We see that the iterations on the price vector $\bs{\lambda}$ generate the corresponding demand response (According to the Federal Energy Regulatory Commission, demand response (DR) is defined as: ``Changes in electric usage by end-use customers from their normal consumption patterns in response to changes in the price of electricity over time, or to incentive payments designed to induce lower electricity use at times of high wholesale market prices or when system reliability is jeopardized.")\cite{drdef,wiki,drdef1,drdef2},
\begin{align}
DR(\bs{\lambda}) = \frac{\partial \bs{u}^{\bs{\lambda}}}{\partial \bs{\lambda}},
\end{align}
where $\bs{u}^{\bs{\lambda}}$ contains the vectors $\bs{u}_i^{\bs{\lambda}}$ for agents $i$ which are consumers. DR is a useful quantity because the social welfare of the grid depends upon it.

Based on the sub-gradient algorithm, we obtain the following price iteration algorithm. Set $k$, the iteration index to $0$. The ISO declares a price vector $\bs{\lambda}^{0}$ (which is chosen arbitrarily, but of course preferably close to the true price vector). 
\begin{itemize}
\item The users $i$ for $i=1,2,...,N$ solve their individual optimal control problems and calculate the $\bs{u}^{\bs{\lambda}(0)}_i$. Then they separately submit their bids $\bs{u}^{\bs{\lambda}(0)}_i$.
\item The ISO then updates the price vector as: $\bs{\lambda}^{k+1} = \bs{\lambda}^{k} - \alpha \left(\sum_i \bs{u}_i\right)$, where $\alpha >0$ is a step size. 
 Increment $k$ by one and go to step $i$.
\end{itemize}

There are several choices for the step size $\alpha$, and several convergence results for the resulting
sub-gradient method~\cite{boyd}.
\section{Bidding with Stochastic Renewables and Demands}\label{stochastic}
In the previous section, the dynamics of the users were assumed to be deterministic, i.e., the exact value of the system state at the next instant was completely determined by~\eqref{dyn}. This might be unrealistic keeping in mind the stochastic nature of renewable energy as well as user demands, etc. 
We begin our discussion with a special case in which the theory can be fully developed. In this case, which we call the Common Completely Observed Case, the sources of stochastic uncertainty are known to all the agents and observed causally by all of them. This could include for example the cloud cover in Denver or wind speed and direction in Brazos County in Texas.
\subsection{Common Observed Randomness Case}\label{CKR}
in which Let $\omega = \omega(1),\omega(2),\ldots,\omega(T)$ be $T$ primitive random variables. They can be independent and identically distributed or Markov. For simplicity, let suppose that each $\omega(t)$ assumes value in a finite set. The state of the $i$-th agent evolves as,
\begin{align*}
X_i(t+1) = f_i^t(X_i(t),U_i(t),\omega(t+1)),
\end{align*}
and it is assumed that each agent observes $\omega$ causally in time, i.e. has access to $\omega(1),\omega(2),\ldots,\omega(t)$ at time $t$. Also everybody knows the probability law $P$ of $\omega$. That is why we call this the Common Observed Randomness Case. The primitive random variables could model the wind speed in Texas or sunlight in Denver, which everybody has access to
causally, and for which they have a dynamic model.

The problem of interest is then to, maximize the utility function
\begin{align}\label{cr}
 & \max \mathbb{E}\left\{\sum_{i=1}^{N}  \sum_{t=1}^{T} F_i(X_i(t))\right\}
\tag{Common Known Randomness Problem}
\end{align}
for the $N$ stochastic dynamic systems
\begin{align}\label{cr1}
X_i(t+1) = f^t_i(X_i(t),\omega(t))
\end{align}
Each $U_i(t)$ is required to be $\mathcal{F}_t$-measurable, where $\mathcal{F}_t = \sigma(\omega(1),\omega(2),\ldots,\omega(t))$ is the sigma-algebra generated by the random variables $\omega(1),\omega(2),\ldots,\omega(t)$.
The inputs $U_i(t)$, for $i=1,2,...,N$ at each $t$ have to satisfy the constraint
\begin{align}\label{cr3}
\sum_i U_i(t) = 0, \mbox{ for each } t = 1,2,\ldots,T.
\end{align}
We consider the following ISO based approach to solving this problem.
Let $\omega^t = \left(\omega(1),\omega(2),\ldots,\omega(t)\right)$ be the past of $\omega $ ($=\omega^{T}$) until time $t$. The ISO announces a price random variable $\lambda(\omega) = \left(\lambda(1,\omega^1),\lambda(2,\omega^2),\ldots,\lambda(T,\omega^T)\right)$ for each $\omega$. 

Note that the price announcement by the ISO is actually a policy announcement.
(Much like the Federal Reserve saying that interest rates will rise if there is a hurricane).
The ISO is saying that if the disturbances $\omega(1),\omega(2),\ldots,\omega(t)$ hit the system by time $t$, then the price will be $\lambda(t+1,\omega^t)$.

Based on this policy announcement, the individual agents also respond with a policy.
Agent $i$ announces a policy $U_i(1,\omega^1),U_i(2,\omega^2),\ldots,U_i(T,\omega^T)$.
The agents determine their policies individually simply by dynamic programming since each knows the
probability law for the stochastic process $\omega$, and their own dynamic system model.

Now we can see that this system is amenable to the same iteration for prices $\lambda(w)$ as before, with the only modification that the iteration process is repeated at each time $t$ to
determine future policies. To elaborate, at each time $t$, the ISO first announces the future price policy as its first iterate,
$\lambda^0(t,\omega^t), \lambda^0(t+1,\omega^{t+1},\ldots,\lambda^0(T,\omega^T)$.
Each agent $i$ then responds with future consumption/generation policy
$U_i(t,\omega^t), U_i(t+1,\omega^{t+1}),\ldots,U_i(T,\omega^T)$.
The ISO computes whether there is a net surplus or deficit of energy at each future time ,
${sum_{i=1}^{N} U_i(t), sum_{i=1}^{N} U_i(t+1),...,sum_{i=1}^{N} U_i(T)}$.
based on this it iterates to produce a new iterate 
$\lambda^1(t,\omega^t), \lambda^1(t+1,\omega^{t+1},\ldots,lambda^1(T,\omega^T)$.
This iteration can be based on a sub-gradient method where the increment is proportional
to the energy surplus/deficit vector.
Then the users again respond with the next iterate of the
future consumption/generation policy
$U_i(t,\omega^t), U_i(t+1,\omega^{t+1}),\ldots,U_i(T,\omega^T)$.
This continues until there is convergence.
 
This is a solution of the problem~\ref{cr} which leads to optimal
utility.
The only issue is complexity, since $\omega$ lies in a huge cardinality set $\vert \Omega \vert$. Thus, we next propose an approximation algorithm.

\textbf{Approximation Algorithm with $k$-step Look-ahead}
At each time $0\leq s\leq T$, the ISO announces the prices $\lambda(s+1),\ldots,\lambda(s+k)$ for the next $k$ time periods, freezing the prices
after $k$ periods.
Iteration then takes place over the $k$-dimensional space, and at each step the iteration tries to achieve energy balance over the next $k$ time periods via bids. The idea is similar to the Model Predictive Control (MPC), so that optimization is performed only for $k$ horizon look-ahead instead of entire $T$ horizon, thus giving us a reduction in complexity. This policy will not approach the optimal policy even as $k\to \infty$ since it is what is called an ``open loop feedback policy". At each time the future price sequence is assumed to be deterministic, not a fully uncertainty state-dependent policy.
\subsection{The Partially Observed Randomness Case}
As opposed to the assumption in the previous section, we consider a more general case where each agent $i$ has a separate ``private" stochastic process  $\omega_i=\omega_i(1),\ldots,\omega_i(T) $ affecting only his system via the equation
\begin{align*}
X_i(t+1) = f_i^t(X_i(t),U_i(t),\omega_i(t+1)),
\end{align*}
The stochastic process $\omega_i$ is not completely observed by the other agents, and only agent $i$ knows the law of process $\omega_i$. The objective function, and the constraints remain the same as in~\ref{cr}. However the assumption that an agent does not have access to the randomness of other agents makes it difficult to achieve the co-operation amongst the agents. 

If the goal is to optimize the utility over all decentralized policies, then the ISO has to know much more about each individual agent's system. It needs to play a more active role so as to induce co-operation amongst the agents. More concretely, the ISO needs to much more about each individual agent and its dynamic system. It needs to know their value of the states $X_i(t)$, utility functions $F_i(\cdot)$ and their dynamics  $f^t_i(X_i(t),U_i(t),\omega_i(t))$  and the probability distributions of their uncertainties $\omega_i$ of each agent $i$. Under this assumption, the ISO can decide the optimal allocations $\bs{U}(t)$ for each $t$, as a function of the state of the system via dynamic programming. This procedure suffers from the curse of dimensionality as the number of users is increased, since it amounts to nothing less than an optimal solution of the decentralized problem.

An optimal decentralized solution, where the solution is itself computed in an iterative decentralized manner to this is an interesting and open problem to pursue.
What we have done above in the Common Completely Observed Randomness case, is demarcated a problem for which the solution is precisely known in principle.

We now present another approach, a relaxation, that also provides very interesting approximation algorithms with much reduced complexity. 

\textbf{Free Storage Relaxation for The Relaxed Partially Observed Randomness Problem}

Let us assume that the ISO has access to a subset of the randomness $\{\omega_i\}_{i=1}^{N}$, which is denoted by $\omega^{ISO}$, knows the law of $\omega^{ISO}$, and assume $\omega^{ISO}$ to be a positive recurrent Markov process. This is the same as assuming that the ISO of a city has knowledge to the weather of the city, or has knowledge of events which might alter the electricity consumption in a big way, and knows the probability laws governing them. However ISO doesn't  have knowledge of the utility functions $F_i$ of the agents, their dynamics $f_i$, nor the entire randomnesses $\omega_i$. Also it is assumed that $\omega^{ISO}$ and its law are known to each agent.
The key idea for producing a tractable approximation is to relax the constraint of energy supply equal to energy consumption at each time $t$, and along each sample path of the stochastic uncertainty, i.e., almost surely. We replace this almost sure equality constraint at each time $t$ by a conditional expectation of the net power being in balance:
\begin{align*}
\limsup_{T\to \infty}\frac{1}{T}\mathbb{E}\left(\sum_{t=1}^{T} \sum_{i=1}^{N} U_i(t)1\left (\omega^{ISO}(t)=w \right)\right)=0,
\end{align*}
where $w$ is any element of the state space of $\omega^{ISO}$. Intuitively it means that the power-balance constraint $\sum_{i=1}^{N} U_i(t)$ is allowed to be violated, however the fluctuation should balance out over time, conditioned on the ISO's observations.

It can be shown that the optimal policy for this case is for the ISO to declare the price at time $t$  based on the value of $\omega^{ISO}(t)$, with users then choosing the quantities $U_i(t)$ based on the value of their state $X_i(t)$, and the process $\omega^{ISO}(t)$. That is, an agent $i$ need not know the value of the state $X_j(t), j\neq i$, nor their utility functions to decide the quantity $U_i(t)$. The analysis and the proof of optimality in the case of large number of agents is analogous to the treatment of multi-armed bandits problems and activity allocation problems \cite{Whi88,Whittle2011Book,Whi80,weber} and uses the technique of large-deviations for Markov process.  We note that the relaxation provides a precise upper bound on utility, which is the utility that can be realized in case there is free storage. It decouples temporal constraints on energy balance.

\textbf{The Limited Lookahead Approach for the Partially Observed Randomness Case}
The Limited Lookahead can also be applied to the Partially Observed case to yield an approach that resembles the Model Predictive Control approach.
We would like to mention in passing that the MPC approach discussed in the Section~\ref{CKR} can be applied in order to develop an approximation algorithm for The Partially Known Randomness Case. 

At each time $t$, the ISO fixes a $k$-step random price vector $\lambda^{0}$ for the next $k$ time instants. This vector will depend only upon the $\omega^{ISO}$ in a causal way. The agents respond to this vector via calculating optimal bids $U_i$  for the future time periods. Then the ISO iterates the price upon receiving the agents' bids. The iterations continue till the changes in the iterates become small enough.

\section{Numerical results}\label{simulations}
We illustrate the above algorithms by a simple example. We start with a deterministic case, followed by a stochastic case. Throughout this section, we assume that the $N$ users are divided into two groups: user $i\in\{1,...,M\}$ act as residential consumers and users $i\in\{M+1,...,N\}$ acts as power suppliers. 
\subsection{Deterministic case}
We first define the state equations. For consumers, $X_{i}(t)$ denotes the room temperature for $i$-th user at time $t$ and $X_{i}(t)$ evolves as, 
\begin{equation*}
X_{i}(t+1)=a_{i}X_{i}(t)+h_{i}-\beta_{i} U_{i}(t),\ i\in\{1,...,M\}
\end{equation*}
where $a_{i}$'s and $\beta_{i}$'s are constant and $h_{i}$ denotes ambient heating.
For suppliers, $X_{i}(t)$ denotes the power production level for the $i$-th user at time $t$, and it evolves as,
\begin{equation*}
X_{i}(t+1)=a_{i}X_{i}(t)+U_{i}(t),\ i\in\{M+1,...,N\}
\end{equation*}
with the $a_{i}$'s being constants.

There are natural constraints associated with the state equations. For consumers, $U_{i}(t)\leq \frac{1}{\beta_{i}}(h_{i}+c_{i})$, where $c_{i}$ is the maximal cooling rate. For suppliers, $U_{i}(t)\leq r_{i}$, where $r_{i}$ is the maximal ramp rate allowed.

We now define the utility functions $F_{i}$. For consumers, let
\begin{equation*}
F_{i}(X_{i}(t))=-\left(X_{i}(t)-\frac{\phi_{1i}+\phi_{2i}}{2}\right)^{2}+m_{i}-\lambda(t)U_{i}(t)
\end{equation*} 
where $[\phi_{1i}$,$\phi_{2i}]$ is $i$-th user's ``comfort bounds'' for temperature, $m_{i}$'s are constant and $\lambda(t)$ is the price.
For suppliers,
\begin{align*}
F_{i}(X_{i}(t))=\lambda(t)X_{i}(t)-&\left(C_{1i}X_{i}^{2}(t)+C_{2i}X_{i}(t)+C_{3i}\right.\nonumber\\&\left.+C_{4i}U_{i}(t)\right).
\end{align*}
where $C_{1i}$, $C_{2i}$, $C_{3i}$ and $C_{4i}$ are cost coefficient for $i$-th user.

In this case, the state equations and constraints are linear, and the objective functions are quadratic, thus we use $QCQP$ (Quadratic Constrained Quadratic Programming) to solve the problem. For simplicity, we let $M=5$, $N=10$, $h_{i}=\beta_{i}=1$, $m_{i}=2$, $a_{i}=1$ and choose $\phi_{1i}$ uniformly from $[20,21]$, $\phi_{2i}$ from $[24,25]$, $r_{i}$ from $[0.5,1.5]$, $C_{1i}$ from $[0.9,1.1]$, $C_{2i}$ from $[0.1,0.3]$, $C_{3i}$ from $[0.5,1.0]$ and $C_{4i}$ from $[0.1,0.5]$. 

\begin{figure}[!t]
\centering
\includegraphics[width=\hsize]{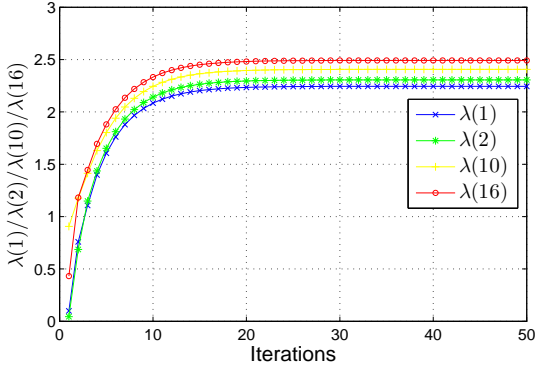}
\caption{Convergence of the price vector for deterministic case}
\label{DRdet}
\end{figure}

Fig. \ref{DRdet} plots the evolution of the price vector and for legibility of display, we only plot $4$ components of $\lambda$. It is easy to see that $\lambda$ converges quite fast, in than 20 iterations.

\begin{figure}[!t]
\centering
\includegraphics[width=\hsize]{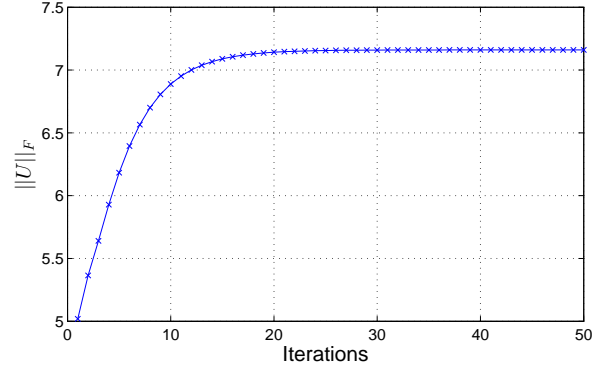}
\caption{Demand Response for deterministic case}
\label{DRdet2}
\end{figure}

Fig. \ref{DRdet2} shows the demand response value as a function of iterations. For the deterministic case, this is simply the norm of matrix $U$, where $U$ is a $M\times T$ matrix with $u_{ij}=U_{i}(j)$. Here we use Frobenius norm defined by: $||U||_{F}=\sqrt{\sum_{i}\sum_{j}u_{ij}^{2}}$. Here we can see that demand response approaches a constant as iteration goes on.

\subsection{Stochastic case}
We adopt the same notation as in the deterministic case, but modify the state equations by adding a random variable influencing the availability of renewable or stochasticity of demand. For consumers,
\begin{equation*}
X_{i}(t+1)=a_{i}X_{i}(t)+h_{i}-\beta_{i} U_{i}(t)+W(t),\ i\in\{1,...,M\}
\end{equation*}
where $W(t)$ is not necessarily i.i.d. because of geographical and temporal correlation.
Similarly, for suppliers,
\begin{equation*}
X_{i}(t+1)=a_{i}X_{i}(t)+U_{i}(t)+V(t),\ i\in\{M+1,...,N\}
\end{equation*}
where $V(t)$ is not necessarily i.i.d. either. 

In our simulation, for simplicity, we let $W(t)$ assume two values drawn uniformly from $[-0.5,0.5]$, each with probability $0.5$; and let $V(t)$ also take two values drawn uniformly from $[-0.2,0.2]$, each with probability $0.5$.

\begin{figure}[!t]
\centering
\includegraphics[width=\hsize]{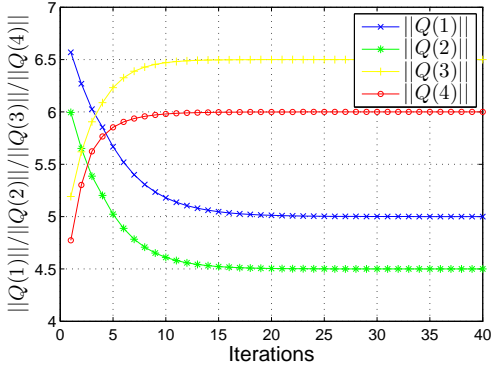}
\caption{Demand Response for stochastic case}
\label{DRsto}
\end{figure}

For each step in the model predictive control approach, the price vector $\lambda$ converges within $20$ iterations, just as it does in the deterministic case. Fig. \ref{DRsto} plots the demand response value as a function of the iterations. Let $Q(t)$ be the vector containing only the $U_{i}$'s for $i\in\{1,...,M\}$. For display purpose, we only plot the first $4$ steps and adopt the $L^{2}$ norm. As the optimization window moves, $||Q||$ converges faster; whether it converges from the above or below depends on the initial value.
\subsection{Comparison with current ISO}
The current ISO sets the price as follows: At time $t$, it assumes that the demand $D(t)$ is given, and based on the previous step's production level $X_{i}(t-1)$, and the marginal cost of each producer, it assigns the production level $X_{i}(t)$ for each producer so as to minimize the total production cost at time $t$. The associated Lagrange multiplier will then be the price at time $t$. We set $a_{i}=3$ for $i\in\{1,...,N\}$, and keep the same value for the other parameters. Each consumer tries to keep $X_{i}(t)=\frac{\phi_{1i}+\phi_{2i}}{2}$ for all $t$, and the resulting $U_{i}(t)$ will be used to calculate the demand input $D(t)$ to the ISO scheme, where $D(t):=\sum_{i}U_{i}(t)$ for $i\in\{1,...,M\}$. 

\begin{figure}[!t]
\centering
\includegraphics[width=\hsize]{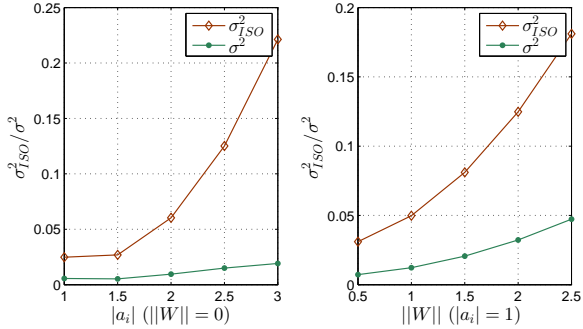}
\caption{Variance of price}
\label{pcomp}
\end{figure}
Fig. \ref{pcomp} summarizes the results. By fixing the uncertainty magnitude $||W||=0$, the figure on the left shows that $\lambda^{ISO}$, which is generated by the ISO scheme, has a bigger variance $\sigma^{2}_{ISO}$ than $\lambda$, which is obtained by our iterative approach. Moreover the difference in variance becomes even larger as $a_{i}$ increases. The figure on the right fixes $|a_{i}|=1$ and plots changes in the variance of price as a function of $||W||$. Similarly in the left figure, the difference in variance increases as $||W||$ increases. 

Next we compare the total utility of the entire system obtained by the two approaches. Notice that in our case the total utility is,
\begin{align*}
u&=\sum_{t=1}^{T}\sum_{i=1}^{N}F_{i}(X_{i}(t))=\\&\sum_{t=1}^{T}\left(\sum_{i\in\{1,...,M\}}-\left(X_{i}(t)-\frac{\phi_{1i}+\phi_{2i}}{2}\right)^{2}+m_{i}\right.\\&\left.-\sum_{i\in\{M+1,...,N\}}C_{1i}X_{i}^{2}(t)+C_{2i}X_{i}(t)+C_{3i}+C_{4i}U_{i}(t)\right)
\end{align*}
as the $\lambda$ terms cancel out. We calculate the total utility incurred by the two schemes; the results are shown in Table \ref{t1}. It can be seen that the total utility obtained by our dynamic iterative approach is roughly three times the total utility obtained by the greedy ISO scheme. (Note that this coincides with the setting of $a_{i}=3$.)

The above considers the case where demand is deterministic. When there is randomness from the demand side, the ISO scheme aims at minimizing the expected production cost at each time $t$. We assume for simplicity that the added noise term $W(t)$ is i.i.d. and takes values $0.5$ and $-0.5$ each with probability $0.5$. Similarly, as in the deterministic case, we calculate the total expected utility incurred by the two schemes and present the results in Table \ref{t2}. It can be seen that the total expected utility obtained by the MPC approach is more than three times the expected utility obtained by the ISO scheme. We conclude that in both deterministic and stochastic cases, our approach provides greater total utility than the current ISO scheme.

\begin{table}[!t]
\centering
\caption{Total utility obtained by ISO scheme and the iterative approach }
\label{t1}
\begin{tabular}{c|c|c}
\hline
\multirow{1}{*}{Name} & \multirow{1}{*}{The iterative approach} & \multirow{1}{*}{ISO scheme} \\
\hline
Value & 427.1932 & 142.8451 \\
\hline
\end{tabular}
\end{table}

\begin{table}[!t]
\centering
\caption{Total expected utility obtained by ISO scheme and the iterative approach }
\label{t2}
\begin{tabular}{c|c|c}
\hline
\multirow{1}{*}{Name} & \multirow{1}{*}{The MPC approach} & \multirow{1}{*}{ISO scheme} \\
\hline
Value & 500.2578 & 159.2198 \\
\hline
\end{tabular}
\end{table}

\begin{figure}[!t]
\centering
\includegraphics[width=\hsize]{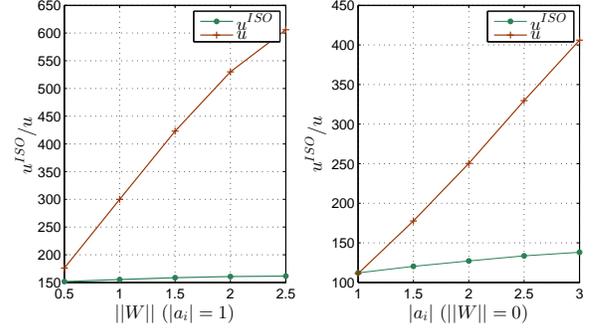}
\caption{Changes in total utility as $|a_{i}|$ or $||W||$ varies}
\label{utsto}
\end{figure}

We also compare the change in total utility obtained as a function of $a_{i}$ for the two schemes. 
We fix $|a_{i}|=1$ and let $W(t)$ be i.i.d. taking values $||W||$ and $-||W||$ with probabilities $0.5$ and $0.5$ respectively. We observe the change in total utility while increasing the noise magnitude $||W||$. The result is shown in the left plot in Fig. \ref{utsto}. It can be seen that as $||W||$ increases, the total utility obtained by the MPC is not a strict linear function of the utility obtained by the ISO scheme. The plot on the right fixes $||W||=0$, and shows that as $|a_{i}|$ increases, the difference in utility obtained increases as well.

\section{Concluding Remarks}
We have formulated the problem of allocating the power demands and generations over the heterogeneous energy consuming or producing agents or prosumers, connected to a smart-grid in a dynamic fashion, both under a deterministic setting and a stochastic setting when there are underlying uncertainties affecting both generation as well as consumption. We have proposed decentralized iterative algorithms to solve this problem. These algorithms work under the assumption of local knowledge, i.e., an agent needs to keep track of only its own randomness and its own system dynamics. We have shown that the ISO can play a central role in inducing co-operation amongst the agents by declaring policies. A possibly significant theoretical contribution is that in the common completely observed randomness case, there is an ISO strategy that achieves social welfare optimality. It incorporates decentralized dynamics where there is no need for agents to be aware of each others' dynamics or states. The only communication from the ISO is price policy and from  the agents their energy consumption or generation in response to the price. We have also proposed more computationally tractable policies for this case. For the case of Partially Observed Randomness case, we have further indicated a relaxation that significantly reduces complexity, as well as an MPC approach that is tractable. Some simulation results showing that the proposed algorithms appear to outperform the current ISO practices in terms of the net social welfare are provided.
\bibliographystyle{IEEEtran}

\bibliography{drref2015mar}

\begin{thebibliography}{10}
\providecommand{\url}[1]{#1}
\csname url@samestyle\endcsname
\providecommand{\newblock}{\relax}
\providecommand{\bibinfo}[2]{#2}
\providecommand{\BIBentrySTDinterwordspacing}{\spaceskip=0pt\relax}
\providecommand{\BIBentryALTinterwordstretchfactor}{4}
\providecommand{\BIBentryALTinterwordspacing}{\spaceskip=\fontdimen2\font plus
\BIBentryALTinterwordstretchfactor\fontdimen3\font minus
  \fontdimen4\font\relax}
\providecommand{\BIBforeignlanguage}[2]{{%
\expandafter\ifx\csname l@#1\endcsname\relax
\typeout{** WARNING: IEEEtran.bst: No hyphenation pattern has been}%
\typeout{** loaded for the language `#1'. Using the pattern for}%
\typeout{** the default language instead.}%
\else
\language=\csname l@#1\endcsname
\fi
#2}}
\providecommand{\BIBdecl}{\relax}
\BIBdecl

\bibitem{Xie2011windintegration}
L.~Xie, P.~M.~S. Carvalho, L.~A. F.~M. Ferreira, J.~Liu, B.~H. Krogh, N.~Popli,
  and M.~D. Ili\'{c}, ``Wind integration in power systems: Operational
  challenges and possible solutions,'' \emph{Proceedings of the IEEE}, vol.~99,
  no.~1, pp. 214--232, Jan. 2011.

\bibitem{Rahimi2010}
F.~Rahimi and A.~Ipakchi, ``Demand response as a market resource under the
  smart grid paradigm,'' \emph{IEEE Transactions on Smart Grid}, vol.~1, no.~1,
  pp. 82--88, Jun. 2010.

\bibitem{An2015}
J.~An, P.~R. Kumar, and L.~Xie, ``On transfer function modeling of price
  responsive demand: An empirical study,'' \emph{Proc. IEEE PES General
  Meeting}, Jul 2015 (to appear).

\bibitem{Callaway2011}
D.~S. Callaway and I.~A. Hiskens, ``Achieving controllability of electric
  loads,'' \emph{Proceedings of the IEEE}, vol.~99, no.~1, pp. 184--199, Jan.
  2011.

\bibitem{Conejo2010}
A.~J. Conejo, J.~M. Morales, and L.~Baringo, ``Real-time demand response
  model,'' \emph{IEEE Transactions on Smart Grid}, vol.~1, no.~3, pp. 236--242,
  Dec. 2010.

\bibitem{Makarov2009}
Y.~V. Makarov, C.~Loutan, J.~Ma, and P.~de~Mello, ``Operational impacts of wind
  generation on {California} power systems,'' \emph{IEEE Transactions on Power
  Systems}, vol.~24, no.~2, pp. 1039--1050, May 2009.

\bibitem{Koch2011}
S.~Koch, J.~L. Mathieu, and D.~S. Callaway, ``Modeling and control of
  aggregated heterogeneous thermostatically controlled loads for ancillary
  services,'' in \emph{Proc. Power System Computation Conference (PSCC)},
  Stockholm, Sweden, 2011, pp. 1--7.

\bibitem{Galus2011}
M.~D. Galus, S.~Koch, and G.~Andersson, ``Provision of load frequency control
  by phevs, controllable loads, and a cogeneration unit,'' \emph{IEEE
  Transactions on Industrial Electronics}, vol.~58, no.~10, pp. 4568--4582,
  Oct. 2011.

\bibitem{Ilic2011framework}
M.~D. Ili\'{c}, J.-Y. Joo, L.~Xie, M.~Prica, and R.~N., ``A decision-making
  framework and simulator for sustainable electric energy systems,'' \emph{IEEE
  Transactions on Sustainable Energy}, vol.~2, no.~1, pp. 37--49, Jan 2011.

\bibitem{Zhu2013}
Q.~Zhu, P.~Sauer, and T.~Basar, ``Value of demand response in the smart grid,''
  in \emph{IEEE Power and Energy Conference at Illinois (PECI)}, Feb. 2013, pp.
  76--82.

\bibitem{Bu2013game}
S.~Bu and F.~R. Yu, ``A game-theoretical scheme in the smart grid with
  demand-side management: Towards a smart cyber-physical power
  infrastructure,'' \emph{IEEE Transactions on Emerging Topics in Computing},
  vol.~1, no.~1, pp. 22--32, Jun. 2013.

\bibitem{Jia2013}
L.~Jia and L.~Tong, ``Day ahead dynamic pricing for demand response in dynamic
  environments,'' \emph{IEEE 52nd Annual Conference on Decision and Control
  (CDC)}, pp. 5608--5613, Dec. 2013.

\bibitem{Song2000optimal}
H.~Song, C.-C. Liu, J.~Lawarree, and R.~W. Dahlgren, ``Optimal electricity
  supply bidding by markov decision process,'' \emph{IEEE Transactions on Power
  Systems}, vol.~15, no.~2, pp. 618--624, May 2000.

\bibitem{Gajjar2003application}
G.~R. Gajjar, S.~A. Khaparde, P.~Nagaraju, and S.~A. Soman, ``Application of
  actor-critic learning algorithm for optimal bidding problem of a genco,''
  \emph{IEEE Transactions on Power Systems}, vol.~18, no.~1, pp. 11--18, Feb.
  2003.

\bibitem{Gao2015optimal}
F.~Gao, G.~B. Sheble, K.~W. Hedman, and C.-N. Yu, ``Optimal bidding strategy
  for gencos based on parametric linear programming considering incomplete
  information,'' \emph{International Journal of Electrical Power \& Energy
  Systems}, vol.~66, pp. 272--279, mar 2015.

\bibitem{Wang2014gametheoretic}
Y.~Wang, W.~Saad, Z.~Han, H.~V. Poor, and T.~Basar, ``A game-theoretic approach
  to energy trading in the smart grid,'' \emph{IEEE Transactions on Smart
  Grid}, vol.~5, no.~3, pp. 1439--1450, May 2014.

\bibitem{Mohsenian2010}
A.~H. Mohsenian-Rad, V.~W.~S. Wong, J.~Jatskevich, R.~Schober, and
  A.~Leon-Garcia, ``Autonomous demand-side management based on game-theoretic
  energy consumption scheduling for the future smart grid,'' \emph{IEEE
  Transactions on Smart Grid}, vol.~1, no.~3, pp. 320--331, Dec. 2010.

\bibitem{drdef}
{Balijepalli, Murthy and Pradhan, Khaparde}, ``Review of demand response under
  smart grid paradigm,'' \emph{IEEE PES Innovative Smart Grid Technologies},
  2011.

\bibitem{wiki}
``Demand response,'' \url{http://en.wikipedia.org/wiki/Demand_response}.

\bibitem{drdef1}
{Albadi, M. H.; E. F. El-Saadany}, ``Demand response in electricity markets: An
  overview,'' \emph{IEEE PESGM}, 2007.

\bibitem{drdef2}
``What is demand response?''
  \url{https://web.archive.org/web/20110819194522/http://www.energydsm.com/demand-response}.

\bibitem{boyd}
``Subgradient methods,''
  \url{https://web.stanford.edu/class/ee392o/subgrad_method.pdf}.

\bibitem{Whi88}
P.~Whittle, ``Restless bandits: activity allocation in a changing world,''
  \emph{Journal of Applied Probability}, pp. 287--298, 1988.

\bibitem{Whittle2011Book}
J.~G.~K. Glazebrook and R.~Weber, \emph{Multi-armed Bandit Allocation
  Indices.}\hskip 1em plus 0.5em minus 0.4em\relax John Wiley \& Sons, 2011.

\bibitem{Whi80}
P.~Whittle, ``Multi-armed bandits and the {G}ittins index,'' \emph{J. R.
  Statist. Soc. B}, vol.~{\bf 42}, pp. 143--149, 1980.

\bibitem{weber}
{R.R. Weber and G. Weiss}, ``On an index policy for restless bandits,''
  \emph{Journal of Applied Probability}, vol.~27, Sep 1990.

\end{thebibliography}
\end{document}